6-16-2021

# Digital Publishing Habits, Perceptions of Open Access Publishing and Other Access Publishing: Across Continents Survey Study

Subaveerapandiyan A
*Regional Institute of Education Mysore*, subaveerapandiyan@gmail.com

Yohapriya K
*Indian School of Business Hyderabad*, yohapriyak@gmail.com

Ghouse Modin Nabeesab Mamdapur
*Synthite Industries Pvt. Ltd*, ghouse@synthite.com

# Digital Publishing Habits, Perceptions of Open Access Publishing and Other Access Publishing: Across Continents Survey Study

**Subaveerapandiyan A**
Professional Assistant
Regional Institute of Education Mysore,
Mysore, Karnataka, INDIA
Email: subaveerapandiyan@gmail.com

**Yohapriya K**
Professional Assistant
Indian School of Business Hyderabad,
Hyderabad, Telangana, INDIA
Email: yohapriyak@gmail.com

**Ghouse Modin Nabeesab Mamdapur**
Scientist-Information & Documentation
NPD & R, Synthite Industries Pvt. Ltd.,
Kolenchery, Ernakulam, Kerala, INDIA
Email: ghouse@synthite.com

**Abstract**

*In this transformative world, changes are happening in all the fields, including scholarly communications are trending in the academic area of publication and access to the resources, especially emerging the wave of open access, open science and open research. The study aims to investigate the digital publishing behaviour of manuscript authors. This study applied a quantitative approach and survey questionnaire method. The researcher collected the data from 251 authors, editors, and peer-reviewers from 45 countries worldwide. The research mainly focuses on the importance, need, and author preference for open access journals. Everyone cannot use and access subscription-based journals; the critical reason is the cost of purchasing a tremendous amount. As an independent researcher, developing countries and other impoverished countries, researchers can give the utmost importance to open access journals. The author also wishes to publish a journal in open access only. The findings reveal that most authors like to publish digital and print in both formats, with chargeless publications. Open access publishing has a vital role with researchers, scholars, and students because accessing the articles is costless. The researcher publishing the manuscript is more important than the quality of the content also important in scholarly publication. Nowadays, open-access peer-reviewed journals are also equal to the paid journals.*

**Introduction**

In earlier days, the traditional system was hiding the information from someone or social communities due to various educational, economic and sociological reasons and the cost of information to publish and disseminate. In this context, many national pioneers, religious & social reformers and other administrative and subject experts had brought many initiatives to make information for all. In order to do that, electronic and digital information has maintained a crucial role while spreading the information to all information seekers without any restrictions. Electronic digital information and communication are easy to access in various forms such as e-books, e-journals, e-reports, e-zine, e-papers and other reference forms. It can be accessed at one's convenience from everywhere and anything through different digital devices and gadgets. The e-Databases, digital libraries, digital archives, digital consortiums, digital repositories, content management portals & getaways, information gateways, news portals and open access directories are functioning as commercial as well as open-access information mediators or aggregators to provide information on open mode or restricted with login Ids. On the flip side numerous academic social networking sites are helping to visualize the scholarly papers such as ResearchGate, Academia, Google Scholar, LinkedIn, nationalized digital projects and innumerable support to effortless search, access, retrieve, preserve and share it for readability and visibility with others. Whatever digital publishing has been creating countless opportunities for all, but it is not error-free. It associated with technological disruptions such as red tape and regulation, content aggregation, trust and transparency, the hunt for viewability, creeping growth of voice tech, rise Gen Z, cookies crumble, protecting digital contents, IPR & copyright ordinance, obsolescence technology, changing documents, different metadata formats, multilingual languages, the behaviour of service providers, low level of user awareness and importantly different opinions about digital or open content quality concern issues are create gaps between the information and its user. Thus, many more minor, major and scholarly publication level research studies have been conducted on digital publishing aspects to analyse the advancements, benefits, service, facilities and facing problems and find some appropriate solution for existing issues. In this perspective, the author considers the same study to assess the digital publishing habits, perceptions of open access publishing, and other access publishing perspectives.

**Review of Literature**

For research, the author searched and browsed the scholarly articles by consulting different full-text, indexing and abstracting core collection databases, digital archives and open access directories. The relevant full-text, abstract and indexing articles, working papers, books, reports, thesis & dissertations scholarly works were collected, reviewed and presented in the following section.

**Ghosh** & **Kumar** (2007) conducted a case study of open access and institutional repositories in developing countries. The research discussed institutional repositories in India. India is one of the leading countries in the open access movement among all the developing countries but fabricating the developed countries recognizes standards of scholarly literature. Still, developing countries face many challenges while accessing scholarly articles and research papers. Most of the peer review journals are costly, and minimal articles are available freely. In developing countries, research publications do not get more audience and attention in the worldwide developed countries. So the Open Access Movement, open access scholarly articles, institutional repositories, and archives make more possible changes to the inaccessibility of the sources.

**Palmer & et al. (2009)** conducted a survey study on academic librarian attitudes about open access. Their analysis suggests that academic libraries and librarians are more entangled in scholarly communication over the work of the institutional repositories. The library should allocate the internal fund and seek the external fund to explore the open-access related work. Academic libraries have to create a separate division and professional position to manage open access and associated scholarly communication projects and issues. Libraries face many challenges to adopting open access, such as legal issues, ethical issues, and scholarly publication pricing. Chief librarians have to take primary responsibility for scholarly communication open access initiatives and library professionals have to conduct the open access campaign and awareness to the users. They finally concluded librarians' involvement, institute financial support, and promoting open access, scholarly publishers support, and funding can make an open access movement possible.

**Creaser (2010)** surveyed the impact of open access and research outputs in the UK. The survey was conducted with academic libraries, and it compared the researcher's practices and perceptions. In this research result, researchers most lack awareness of their institutional policies, institutional repositories, open access, and researcher or mistrust of open access publication. Simultaneously with a degree of obliviousness about open access and the prominent role of institutional repositories. Some significant findings are that 10% of institute respondents were not aware of how to pay for publishing open access journals, they do not get funds for open access publishing. The responding institutes typically encourage self-archiving but researchers lack awareness of this. The library professionals notify the majority of the 93% of respondent's institutes to promote self-archiving in their institute repository, 92% of respondents libraries are taken responsibility for the institutional repository. 70% of researchers and academicians did not know where the institute has a repository. The crucial factor is 74% of respondents like to publish open access because the information disseminates quickly and widely.

**Pagliaro (2021)** discussed open access publishing in chemistry; in earlier years, chemistry research scholars adopted open access publishing without restriction. Publishing manuscripts highly improves visibility and without any fee. He also discussed how they adopted earlier open access, but nowadays, it's faded away. He did the article investigation between 2009 and 2015 in basic science; in 2016 chemistry was lowest in open access articles. The main reason is the lack of funds to publish an open-access article.

**Schilhan & et al. (2021)** discussed how to present the scholarly article effectively to the researchers. With the help of academic search engines and databases. SEO is already an existing concept used for frequent marketing purposes, and it boosts up the retrieval of the document in any kind of format. ASEO is specially used for scholarly academic search texts and findings-related queries. They give some most critical points: choosing keywords wisely, a clear title, and rich metadata. ASEO is highly helpful to the authors and publishers who know the altmetrics, downloads, viewers, readers, citations, and so on. Depending on the search engine result will vary because of the optimization techniques. Some search engines give the search result related factors such as latest articles, books, and posts; highly cited or viewed articles. Some ideas also discussed do's and do not in scholarly communications. Do the meaningful title, necessary phrase, word upfront, imagine in explorable terms, make it a sufficient title, use thesauri, narrow vs broad terms, use the singular form, write the perspective of information seeker, short sentences, use synonymous, and repeat the keywords. Do not avoid special characters, signs and ambiguous titles.

**Significant of the study are**
1. To identify the author's journal publication preferences
2. To understand the expectation of authors
3. To know the way of visualizing the journals publicly
4. To find out the present importance of publishing paper
5. To compare the paid vs free publication

**Materials and Methods**
The researcher used a quantitative research design. A well-structured questionnaire was distributed to the authors and researchers who previously published their research papers in Scopus and other peer-reviewed journals. Respondents email IDs collected from various respective journal websites worldwide. The closed-ended questionnaire comprises 22 questions and 21 questions made compulsory, and 1 question was optional and the study samples consisted of 251 peer-review published authors from 45 countries and across the world. A disproportionate stratified sampling technique was used to collect the data. The statistical techniques were used for the analysis of data in Excel and the analysed data has been presented in the following tables and figures.

**Data Analysis and Interpretation**

Table 1 shows the number of respondents according to continents of the world. Totally 251 peer-review published authors participated in this study, out of which 44.6% authors from Asia, followed by Africa 10%, Europe 24.3%, North America 10.4%, South America 8.4% and Australia/Oceania 2.4%. The table data clearly shows that the highest number of respondents are from Asian countries with 44.6% and the least respondents from Australia, only 2.4%.

**Table 1. Distribution of respondents according to Continents**

| Continents | Respondents | Percentage |
|---|---|---|
| Asia | 112 | 44.6 |
| Africa | 25 | 10 |
| Europe | 61 | 24.3 |
| North America | 26 | 10.4 |
| South America | 21 | 8.4 |
| Australia/Oceania | 6 | 2.4 |

Further, Table 2 revealed that the qualification wise distribution of respondents. It is found that the highest 71.3% of participated authors in the study had obtained PhD in their respective subject field and second-highest 19.9% of them are Post-Graduate holders and then remaining are 7.2% are holding M.Phil. Degree and a few more 1.6% authors had an under-graduation course in their respective field.

**Table 2. Distribution of respondents according to qualification**

| Qualification | Respondents | Percentage |
|---|---|---|
| **PhD.** | 179 | 71.3 |
| **M.Phil** | 18 | 7.2 |
| **PG** | 50 | 19.9 |
| **UG** | 4 | 1.6 |

It is found from Table 3 highlights the research productivity of the authors in journal publications. It is revealed that the highest 34.7% of authors had published more than 25 research publications which indexed in the Scopus core collection citation databases and secondly 6-10 articles were published by 25.1% and then 1-5 articles from 22.3% of authors.

**Table 3. Number of Articles Published by the Respondents**

| No. of Articles Published | Respondents | Percentage |
|---|---|---|
| 1-5 | 56 | 22.3 |
| 6-10 | 63 | 25.1 |
| 11-15 | 10 | 4 |
| 16-25 | 35 | 13.9 |
| Above 25 | 87 | 34.7 |

Table 4 indicates authors usually like to choose a research methodology for their study. The below table shows us the highest 37.1% of respondents prefer a mixed method of research. In contrast, qualitative 17.5%, quantitative 13.9%, 5.2% and use the experimental, depending on the articles 2.4% and 22.3% of authors are other un-specified research methods during the study.

**Table 4. Most Preferred Research Methods by Authors**

| Methodology Preference | Respondents | Percentage |
|---|---|---|
| Quantitative | 35 | 13.9 |
| Qualitative | 44 | 17.5 |
| Empirical | 4 | 1.6 |
| Mixed method | 93 | 37.1 |
| Experimental | 13 | 5.2 |
| Depends on the Article | 6 | 2.4 |
| Other | 56 | 22.3 |

Table 5 reveals the publishing opinions of the author on research contributions to newly established journals. It's evident from the above below table reveals that the highest 42.6% of respondents felt that they are might or might not consider publishing their research papers in newly established journals, followed by 31.5% of them felt as definitely consider and would not consider 19.1%, and 6.8% of them are not willing to publish their contribution due to unspecified other reasons.

**Table 5. Author's perceptions on contributing to newly established journals**

| Contributing opinion of a newly established journal | Respondents | Percentage |
|---|---|---|
| Definitely consider | 79 | 31.5 |
| Might or might not consider | 107 | 42.6 |
| Would not consider | 48 | 19.1 |
| Other reasons | 17 | 6.8 |

It is evident from Table 6 and Figure 1 examines the type of document format preferred by the authors to publishing their research papers; in this context, the study found that the highest 52.6% of authors are showing interest to publish their article both in electronic and print format and second-highest 45.8% of them preferred only electronic version of journals, may they feel if they publish their papers in electronic version and then their paper gets more citation and getting more refer then the print format. Thus, only a few 1.6% prefer the print format of journals.

**Table 6. Type of document format preferred by the authors**

| Publication Preference | Respondents | Percentage |
|---|---|---|
| Electronic | 115 | 45.8 |
| Print | 4 | 1.6 |
| Both | 132 | 52.6 |

**Figure 1. Type of document format preferred by the authors**

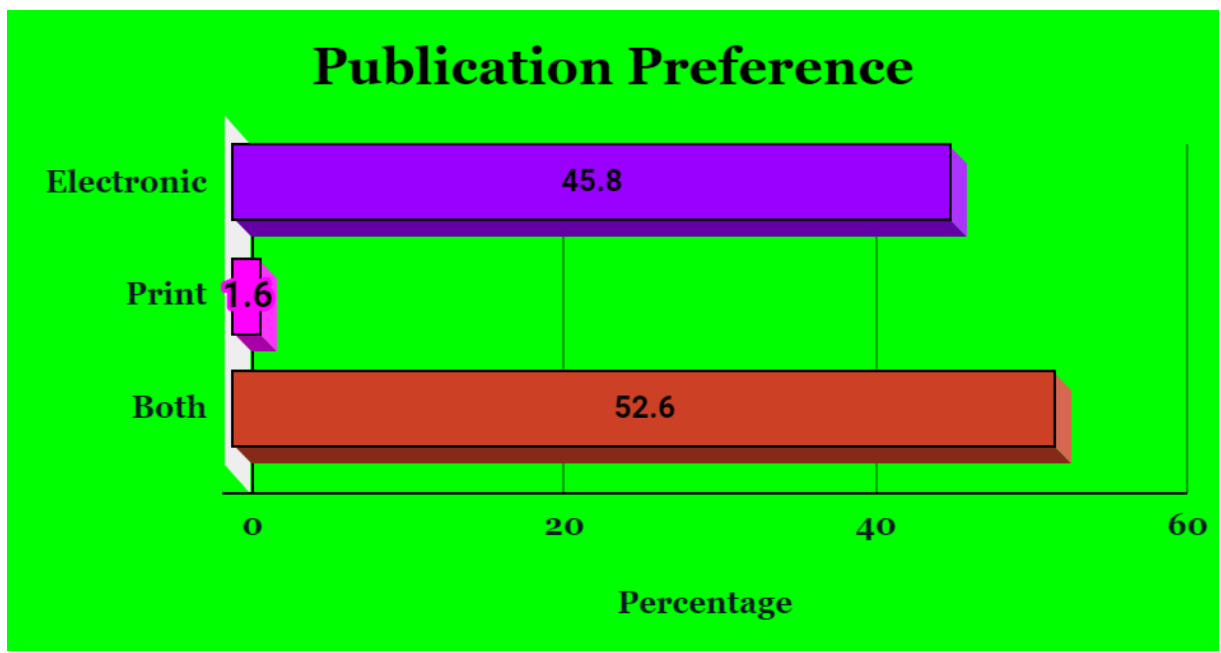

Below Table 7 and figure 2 pointed out significant reasons for choosing digital publications of journal articles. It is shown that the highest 89.7% and 74.5% of authors were felt with that accessibility to all readers (including print disabled), Covering a large base of readers and Higher acceptance from readers issues are most influenced on the authors to give first priority to choose digital publications to publish their paper as well as refer to prepare the papers and few more of the authors are disagree with all two statements.

**Table 7. Reasons for selection of digital publications**

| Reasons for choosing the Digital Publication | Agree | Neutral | Disagree | Total |
|---|---|---|---|---|
| **Higher acceptance from readers** | 187 (74.5%) | 51 (20.3%) | 13 (5.2%) | 251 (100%) |
| **Covering a large base of readers** | 225 (89.7%) | 26 (10.3%) | 0 (0%) | 251 (100%) |
| **Accessibility to all readers (including print disabled)** | 225 (89.7%) | 20 (8%) | 6 (2.3%) | 251 (100%) |

**Figure 2. Reasons for selection of digital publications**

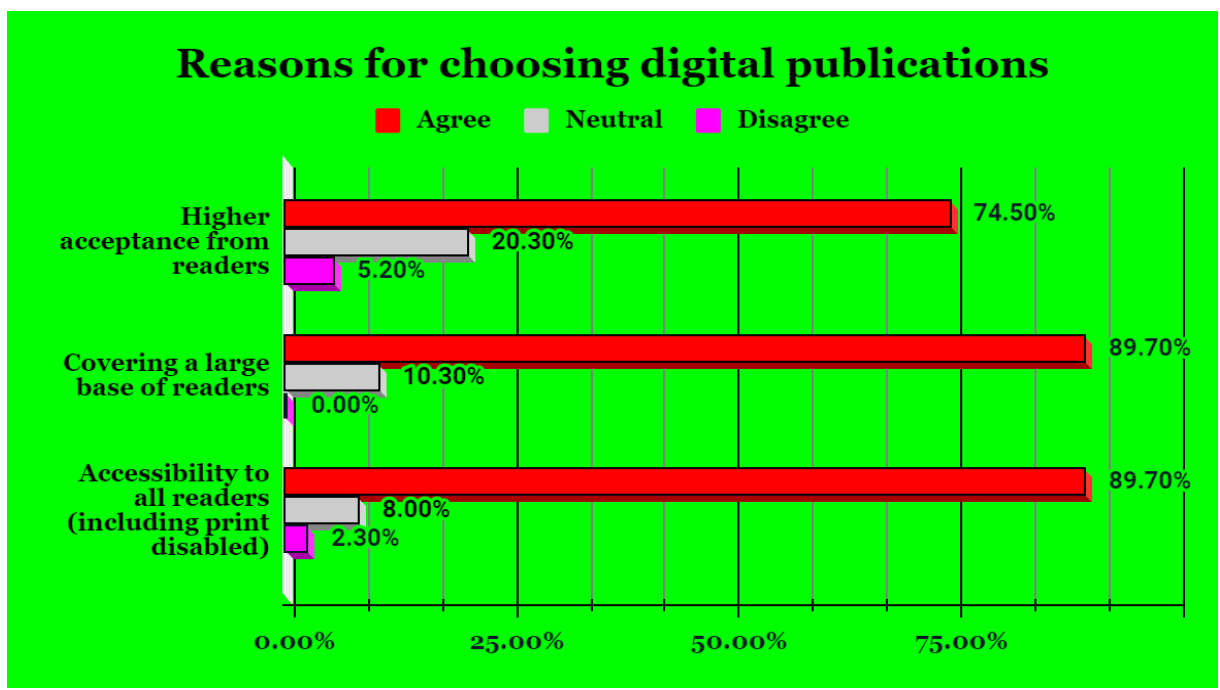

It is found from above Table 8 and figure 3 analysing the awareness of journal identifiers /indicators, out of 251 total respondents 73.7%, 72.5% & 71.8% of respondents are expressed they have extreme awareness of Digital Object Identifier (DOI), impact factors and ISBN. On average, 40.28% expressed that they are moderately aware of all Journals identifiers such as DOI, P-ISSN, E-ISSN, ISBN, PII, EAN, and impact factors. Also, the study found the highest 36.2% & 22.7% of respondents felt that they were *'not at all aware'* of the EAN & PII journal identifiers.

**Table 8. Awareness of Journals identifiers by the authors**

| Identifiers/ Impact Factors | Extremely Aware | Moderately Aware | Somewhat Aware | Slightly Aware | Not at all Aware | Total |
|---|---|---|---|---|---|---|
| **DOI** | 185 (73.7%) | 39 (15.5%) | 22 (8.8%) | 5 (2%) | 0 (0%) | 251 (100%) |
| **P-ISSN** | 132 (52.6%) | 57 (22.7%) | 25 (10%) | 15 (6%) | 22 (8.7%) | 251 (100%) |
| **E-ISSN** | 153 (61%) | 46 (18.3%) | 23 (9.1%) | 12 (4.8%) | 17 (6.8%) | 251 (100%) |
| **ISBN** | 180 (71.8%) | 41 (16.3%) | 18 (7.1%) | 10 (4%) | 2 (0.8%) | 251 (100%) |
| **PII** | 68 (27%) | 45 (18%) | 54 (21.5%) | 27 (10.8%) | 57 (22.7%) | 251 (100%) |
| **EAN** | 29 (11.6%) | 51 (20.3%) | 46 (18.3%) | 34 (13.6%) | 91 (36.2%) | 251 (100%) |
| **Impact Factors** | 182 (72.5%) | 48 (19.1%) | 12 (4.8%) | 7 (2.8%) | 2 (0.8%) | 251 (100%) |

**Figure 3. Awareness of Journals identifiers by the authors**

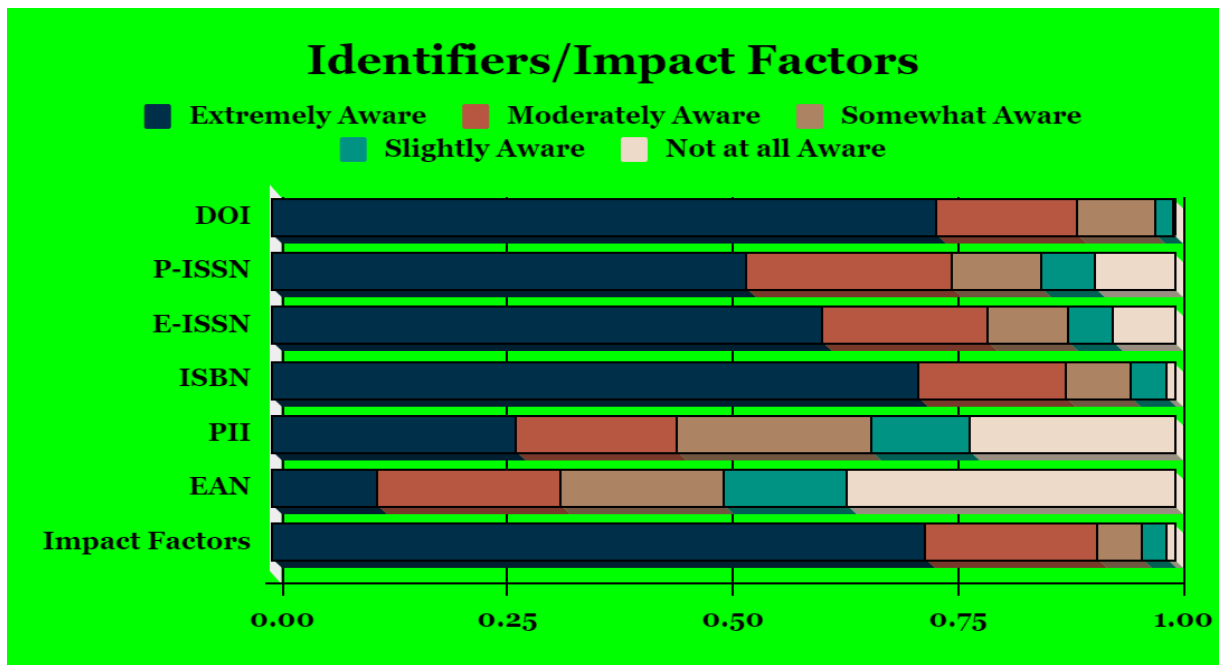

Table 9 and figure 4 investigates the journal publishing ethical awareness about journals, books and other research documents. It is observed from the study, the highest 79.7%, 64.1% & 58.6% of the respondents pointed out that they are extremely aware of plagiarism, publication ethics and copyright. On average, 52.5% of respondents feel that they have *'moderately aware'* of the publication ethics, predatory journals, open access policy, copyrights, creative commons, and plagiarism. Only a few more are expressed as *'not at all aware'* about all six ethical components.

Table 9. Ethical Awareness about Journals, Books, and Others

| Journal Publication Awareness | Extremely Aware | Moderately Aware | Somewhat Aware | Slightly Aware | Not at all Aware | Total |
|---|---|---|---|---|---|---|
| Publication Ethics | 161 (64.1%) | 46 (18.3%) | 18 (7.2%) | 22 (8.8%) | 4 (1.6%) | 251 (100%) |
| Predatory Journals | 140 (55.8%) | 40 (16%) | 39 (15.6%) | 16 (6.3%) | 16 (6.3%) | 251 (100%) |
| Open Access Policy | 135 (53.8%) | 70 (28%) | 23 (9%) | 17 (6.8%) | 6 (2.4%) | 251 (100%) |
| Copyrights | 147 (58.6%) | 64 (25.5%) | 24 (9.6%) | 9 (3.6%) | 7 (2.7%) | 251 (100%) |
| Creative Commons | 93 (37%) | 71 (28.3%) | 41 (16.3%) | 20 (8%) | 26 (10.4%) | 251 (100%) |
| Plagiarism | 200 (79.7%) | 24 (9.6%) | 10 (4%) | 9 (3.6%) | 8 (3.1%) | 251 (100%) |

Figure 4. Ethical Awareness about Journals, Books, and Others

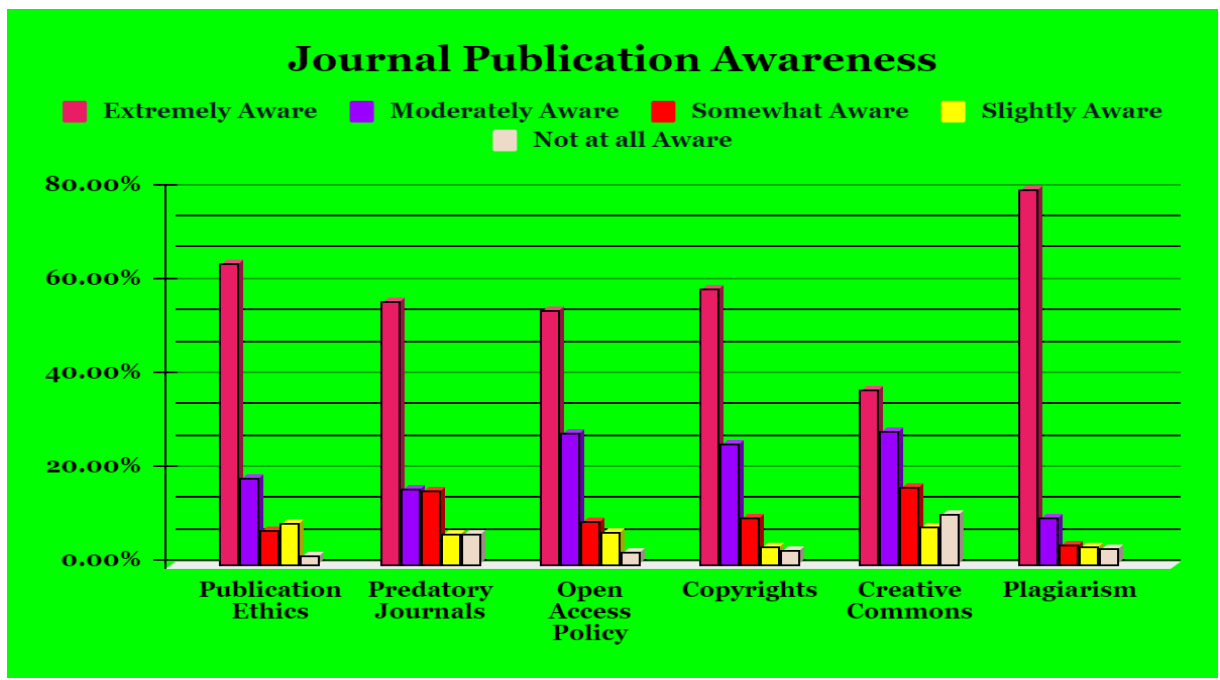

It is found below Table 10 and the figure 5 examines the author's perception of open access benefits and advantages provided for researchers. Results seen from the table indicate that most

65.7%, 64.6%, 49.8% and 45.4% of the authors respectively expressed that they have *'strongly agreed'* with visibility and usage of research, increased citation and use, faster impact and enhancing the research process. Also, a few more on average 1.8% of the authors felt as *'strongly disagreed'* with all research variables as mentioned in the table.

**Table 10. Author perception on Open access benefits**

| Open Access Benefits | Strongly Agree | Agree | Neutral | Disagree | Strongly Disagree | Total |
|---|---|---|---|---|---|---|
| Enhancing the research process | 114 (45.4%) | 87 (34.7%) | 37 (14.7%) | 7 (2.8%) | 6 (2.4%) | 251 (100%) |
| Visibility and usage of research | 165 (65.7%) | 62 (24.7%) | 16 (6.4%) | 3 (1.2%) | 5 (2%) | 251 (100%) |
| Increased citation and usage | 162 (64.6%) | 59 (23.5%) | 21 (8.3%) | 5 (2%) | 4 (1.6%) | 251 (100%) |
| Faster impact | 125 (49.8%) | 80 (31.9%) | 32 (12.8%) | 11 (4.3%) | 3 (1.2%) | 251 (100%) |

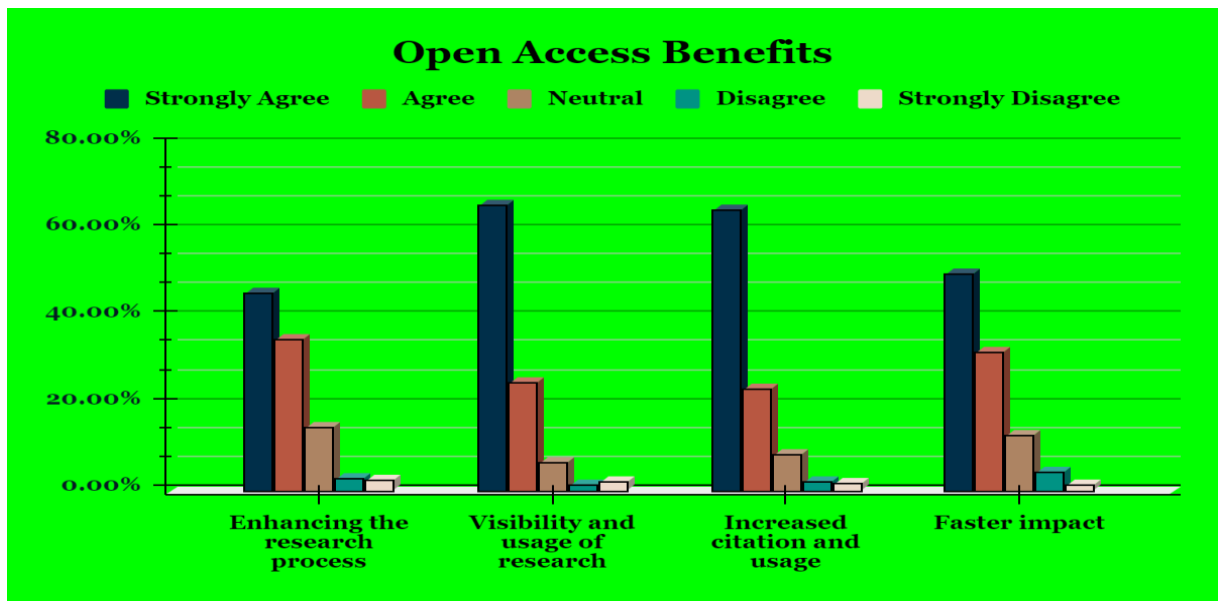

It is found from Table 11 has identified the author's perception of the goodness of the Google scholar while used for citation. Out of 251 total respondents, 67.3% felt as good about the 'Google Scholar' research assistance while searching and retrieving articles sources for research and other aspects and increasing the author's citation index.

**Table 11. Author perception on the goodness of the Google Scholar**

| Google Scholar is a Good Indicator | Respondents | Percentage |
|---|---|---|
| Yes | 169 | 67.3 |
| No | 60 | 23.9 |
| **I am not sure** | 22 | 8.8 |
| **Total** | 251 | 100 |

Table 12 describes open access articles submitting and publishing preferences. The study results showed that most 44.2% of authors were coated that they have given high priority to publishing their articles in open access journals, followed by 29.5% of them giving medium priority and only a few 4.4% of authors are not given any priority to publish their paper in open access mode.

**Table 12. Priority Level Open Access Articles publishing by the authors**

| Priority of Open Access | Respondents | Percentage |
|---|---|---|
| Not a priority | 11 | 4.4 |
| Low priority | 14 | 5.6 |
| Medium priority | 74 | 29.5 |
| High priority | 111 | 44.2 |
| Essential | 41 | 16.3 |
| Total | 251 | 100 |

Table 13 examines the factor influence while selecting a journal by the author to publish research publications. The result from the table indicates that the highest 71.00% of respondents expressed that the Journal reputation in the field is the most impact on the select journal for publishing the papers, followed by 67.7% of them felt that Journal impact factor and third highest 58.9% responses go to Area of specialization and Publication speed has impacted to 48.4% of respondents during the study.

**Table 13. Factors influence during the selection of a journal for publishing**

| Way to choose a journal | Respondents | Percentage |
|---|---|---|
| Journal impact factor | 171 | 67.7 |
| Journal reputation in the field | 176 | 71 |
| Publication speed | 120 | 48.4 |
| Acceptance/rejection rate | 62 | 25 |
| Open access fee | 87 | 33.9 |
| Area of specialization | 149 | 58.9 |
| Established journal | 108 | 43.5 |
| Number of issues per year | 70 | 28.2 |
| Others | 18 | 7.2 |

Following Table 14, analyze the channels or ways used to select the most suitable or appropriate journals to publish their research contributions. The study results indicate that the majority of 68.5% & 65.3% of authors are selected journals based on the journal indexed in well-known citation databases and its aims or scope. Then, 53.2% of respondents considered the impact factor of journals, the free of cost of publishing journals preferred by 47.6% of respondents and 41.1% of them liked the peer-reviewing process journals.

Table 14. Channels of while select appropriate journal to publishing

| Way of finding appropriate Journal | Respondents | Percentage |
|---|---|---|
| Indexed | 173 | 68.5 |
| Impact factor | 137 | 53.2 |
| Aims/scope of the journals | 163 | 65.3 |
| Peer-review process | 102 | 41.1 |
| Audience/readership of the journal | 86 | 34.7 |
| Open access options | 80 | 32.3 |
| Free of cost for publishing | 121 | 47.6 |

Further, in Table 15, the study has extended to analyze the personal/author profile ID's and accounts in SNS created by the respondents. Out of 251 total respondents, the highest 87.2%, 86%, 84.4%, 71% and 69.8% of respondents respectively had created personal/author profile ID' and accounts in Google Scholar, ResearchGate, ORCID, Academia and Scopus open-access databases and directories. They might have created this profile ID' and accounts for educational, teaching, research and other networking or collaborative purposes and used it to showcase their research contributions for further reference and use.

Table 15. Personal/Author profile ID's and accounts in SNS

| Researcher ID/ Account in SNS | Yes | No | Total |
|---|---|---|---|
| ORCID | 212 (84.4%) | 39 (15.6%) | 251 (100%) |
| Google Scholar | 219 (87.2%) | 32 (12.8%) | 251 (100%) |
| Scopus | 175 (69.8%) | 76 (30.2%) | 251 (100%) |
| Microsoft | 84 (33.4%) | 167 (66.6%) | 251 (100%) |
| Researchgate | 216 (86%) | 35 (14%) | 251 (100%) |
| Academia | 178 (71%) | 73 (29%) | 251 (100%) |

It is found from Table 16 shows that 52.6% & 32.3% of respondents are scaled as *'strongly agreed'* & *'agreed'* with that, the high-impact journals can improve visibility of research contents and content modulators, it can reduce quantitative research and increase qualitative research in all research areas and few 4.3% of respondents are given a negative opinion on improves the visibility of research.

Table 16. High impact journal while improves visibility of research

| High impact journals improve the visibility | Respondents | Percentage |
|---|---|---|
| Strongly Agree | 132 | 52.6 |
| Agree | 81 | 32.3 |
| Neutral | 26 | 10.4 |
| Disagree | 11 | 4.4 |
| Strongly Disagree | 1 | 0.4 |
| Total | 251 | 100 |

Further, the study finds ways and possibilities to increase visibility in research. Out of 251 respondents, 69% of respondents were chosen to archive research papers in institutional repositories and subject repositories, 67% of respondents gave their opinion on citing the article whenever appropriate, as shown in Table 17.

Table 17. Increase visibility in research

| Way of increasing visibility of journal paper | Respondents | Percentage |
|---|---|---|
| Cite the paper whenever appropriate | 168 | 67 |
| Post preprint/postprint on arXiv/similar repository | 52 | 20.7 |
| Send announcements out on an appropriate email list | 40 | 20 |
| Add the title to the CV on your home page under "Recent Publications", with a link | 121 | 48.2 |
| post announcements about the article on various social media sites like LinkedIn, Twitter, Facebook, and a dozen others. | 93 | 37 |
| Archive your papers in institutional or subject repositories (e.g. SSRN, Mendeley, ResearchGate, Academia etc.). | 173 | 69 |
| Other | 12 | 4.8 |

The study results had analyzed the most preferred to used reference management software's by the respondents presented in Table 18. The study results showed that majority 41.8% of respondents are most preferred to use the Mendeley citation/reference management software to organize their cited reference materials as different reference manuals and citation standards and present content in a unique manner and the EndNote is considered as second most used software then the Zotero with 10.4% and remaining 21.1% of respondents do not use any reference or citation management software.

**Table 18. Reference Management Software's preferred to use by the Author's**

| Citation Management Software Usage | Respondents | Percentage |
|---|---|---|
| Zotero | 26 | 10.4 |
| Mendeley | 105 | 41.8 |
| EasyBib | 10 | 4 |
| EndNote | 57 | 22.7 |
| None of these | 53 | 21.1 |
| Total | 251 | 100 |

The following Table 19 determined the nature of information content and search strategies used by the respondents. It is found highest 73.70% of respondents felt that title is their first choice to search, browse, use and analyze the research contents, then 68.1% of them are would like preferred abstract and 50.6% of respondents most prefer and refer the methodology parts & research design, 42.6% of them are noticed that the most used recent publications and most cited research are 31.00%. Only a few 6.3% of respondents are looking for information author designation.

**Table 19. Nature of information content and search strategies used by the authors**

| Criteria for choosing reference articles | Respondents | Percentage |
|---|---|---|
| Title | 185 | 73.7 |
| Author affiliation | 47 | 18.8 |
| Author designation | 16 | 6.3 |
| Famous personality | 28 | 11.1 |
| Abstract | 171 | 68.1 |
| Methodology | 127 | 50.6 |
| Recent publications | 107 | 42.6 |
| Most cited | 78 | 31 |
| Most downloaded | 24 | 9.5 |
| Open access | 63 | 25 |
| Other | 12 | 4.8 |

Table 20 presents the nature or types of research publications preferred or referred to by the respondents. The table showed that 65.7% of the highest respondents preferred to send their research papers for free publications, followed by 28.7% of respondents who were ready to publish paid and free both ways. Due to the urgency of research requirements, low-quality content and just educational benefits, 5.6% of respondents would like to publish their paper by paid publications.

**Table 20. Nature of publications preferred by the authors**

| Publication preference | Respondents | Percentage |
|---|---|---|
| Paid publications | 14 | 5.6 |
| Free publications | 165 | 65.7 |
| Both | 72 | 28.7 |
| Total | 251 | 100 |

An interesting fact found from Table 21 shows that the highest 81.1% of participating respondents expressed they have served as editor, associate editor, member editor and editor to manuscript sent registered publications. Thus, to a great extent, respondents are well aware of different research ethics and publications components.

**Table 21. Authors served as Peer Reviewer to Journal Publications**

| Peer-reviewer/Editor | Respondents | Percentage |
|---|---|---|
| Yes | 184 | 81.1 |
| No | 42 | 18.5 |
| Other | 1 | 0.4 |
| Total | 227 | 100 |

**Findings of the study**

- It enlightens that 71.3% of participating authors in the study had obtained a PhD in their respective subject field.
- It determined that the highest 34.7% of authors had published more than 25 research publications indexed in the Scopus core collection citation databases.
- It is shown that the highest 42.6% of respondents felt that they might or might not consider publishing their research papers in newly established journals.

- It is found that the highest 52.6% of authors are showing interest to publish their paper both in electronic and print format.
- It shows that the highest 89.7% prefers to use digital publications because it is available to all readers (including print disabled) and covers a large base of readers.
- It is analysed on average 40.28% of respondents are expressed that they have moderately aware of all Journals identifiers such as DOI, P-ISSN, E-ISSN, ISBN, PII, EAN and impact factors.
- It is found that the highest 79.7%, 64.1% & 58.6% of the respondents pointed out that they are extremely aware of plagiarism, publication ethics and copyright.
- The study results showed most of 65.7%, 64.6%, 49.8% and 45.4% of the authors respectively expressed that they have *'strongly agreed'* with visibility and usage of research, increased citation and usage, faster impact and enhancing the research process.
- It is noticed that the majority 67.3% of respondents felt good about the 'Google Scholar' assistance while searching and retrieving article sources for research.
- It is analysed the most 44.2% of authors were coated that they have given high priority to publishing their articles in open access journals.
- The study found that the highest 71.00% of respondents expressed that the Journal reputation in the field has the most impact on selecting journals for publishing the papers.
- It is found that the majority of 68.5% of authors are selected journals based on the journal indexed in well-known citation databases.
- It is identified that 87.2%, 86%, 84.4%, 71% and 69.8% of respondents respectively had created personal/author profile ID' and accounts in Google Scholar, ResearchGate, ORCID, Academia and Scopus and used for different purposes.
- It highlights that 52.6% & 32.3% of respondents are scaled as *'strongly agreed'* & *'agreed'* with that the high-impact journals can be improved visibility research contents and content modulators.
- The study found that 41.8% of respondents most preferred to use the Mendeley citation/reference management software to organize their cited reference materials and uniquely present them.
- It highlights that 73.70% of respondents felt that title is their first choice to search, browse, use and analyze the research content.

- The study noticed that 81.1% of participants expressed they have served as editor, associate editor, member editor and editor to manuscript sent registered publications.

**Conclusion**

Nowadays, information technology has been creating more opportunities and challenges for both information generators and information seekers. It made provision that anybody can express and produce their views in traditional and digital, virtual, and electronic modes. Thus, now information is generated and available in the conventional form and digital or electronic. In earlier days, electronic information was also not easily available for information seekers due to the IPR, copyright issues, and most electronic resources available from a commercial perspective. In this case, the open access movement has come to exist to provide information to all by breaking all traditional walls and commercial restrictions. Therefore, the researcher attempted to examine the authors or information generators perception of digital and open-access information. The analysed study results have been drawn and presented based on the observations.